\documentclass[useAMS,usenatbib]{mn2e}
\usepackage{amssymb,amsmath}
\usepackage{graphicx}
\title[$M_{BH}$ and $R_{BLR}$ of dbp emitters]
{The sizes of BLRs and BH masses of double-peaked broad low-ionization 
emission line objects}

\author[Zhang et al.]
       {Xue-Guang Zhang$^{1,2}$\thanks{xguang@astroscu.unam.mx}, 
        Dultzin-Hacyan D.$^1$,
        Ting-Gui Wang$^2$ \\     
       $^1$Instituto de Astronomia, Universidad Nacional Autonoma de
                 Mexico, Apdo Postal 70-264, Mexico D. F. 04510, Mexico \\
       $^2$Center for Astrophysics, Department of astronomy and Applied
                 Physics, University of Science and Technology of China, \\
                 Hefei, Anhui, P.R.China}

\date{}

\pagerange{\pageref{firstpage}--\pageref{lastpage}} \pubyear{2006}
\def\LaTeX{L\kern-.36em\raise.3ex\hbox{a}\kern-.15em
    T\kern-.1667em\lower.7ex\hbox{E}\kern-.125emX}

\begin{document}
\label{firstpage}

\maketitle

\begin{abstract}
  In this paper, the sizes of the BLRs and BH masses of DouBle-Peaked 
broad low-ionization emission line emitters (dbp emitters)  are 
compared using different methods: virial BH masses vs BH masses from stellar 
velocity dispersions, the size of BLRs from the continuum luminosity vs 
the size of BLRs from the accretion disk model. 
First, the virial BH masses of dbp emitters 
estimated by the continumm luminosity and line width of broad H$\beta$ are 
about six times 
(a much larger value, if including another dbp emitters, of which 
the stellar velocity dispersions are traced by the line widths of 
narrow emission lines) larger than the BH masses estimated from the relation 
$M_{BH} - \sigma$ which is a more accurate relation to estimate BH masses. 
Second, the sizes of the BLRs of dbp emitters estimated by the empirical 
relation of $R_{BLR} - L_{5100\AA}$ are about three times 
(a much larger value, if including another dbp emitters, of which the 
stellar velocity dispersions are traced by the line widths of 
narrow emission lines) larger than the mean flux-weighted sizes of BLRs of 
dbp emitters estimated by the accretion disk model. 
The higher electron density of BLRs of dbp emitters would be the
main reason which leads to smaller size of BLRs than the predicted value from 
the continuum luminosity.
\end{abstract}


\section{Introduction}

   How to measure the black hole (BH) masses of Active Galactic 
Nuclei (AGN) is an intriguing research subject. The masses of the central 
black hole in nearby galaxies can be estimated by kinematic studies of 
nuclear gas disks (\citealt{har94}, Tsvetanov et al. 1998, 
Ford et al. 1994, Gebhardt et al. 2000b) or dynamical 
studies of stars in galactic centers (Gebhardt \& 
Richstone 2000, Bower et al. 2000), using high spatial resolution data 
(\citealt{geb03}). However, because of higher luminosity and larger distance, 
these two methods are noneffictive for AGN. Thus, there are many efforts to 
find effective ways to estimate the masses of central black hole of AGN. 
The most successful and convenient way is under the assumption of 
virialization, $M_{BH}\sim\upsilon^2R_{BLR}$. The size of Broad Emission 
Line Regions (BLRs, $R_{BLR}$) can be estimated through the empirical relation 
between continuum luminosity and the size of BLRs (Kaspi et al. 2000, 
2005, Peterson et al. 2004, Wandel et al. 1999), according 
to the results of reverberation mapping technique (Blandford \& Mckee 1982; 
Peterson 1993; Netzer \& Peterson 1997). 
Throughout this paper we shall use the term 'size of BLRs' as 
equivalent of distance of the BLRs to the central source.
The velocity $\upsilon$ can be 
estimated by the line width of broad low-ionization emission lines coming 
from BLRs, assuming motions of the BLRs clouds are gravitationally dominated 
by the central mass of the host galaxy (Gaskell 1988, Wandel et al. 1999, 
Peterson \& Wandel 1999, Gaskell 1996). Using these methods, BH masses have 
been determined for a large number of AGN. From BH masses, 
other fundamental parameters of AGN, such as the 
dimensionless accretion rate $\dot{m}$, can be estimated. Thus, the nature 
of AGN can be understood better.

  The empirical relation between $R_{BLR}$ and continuum luminosity 
$L_{5100\AA}$ has been studied for several years, since it was found by Kaspi 
et al. (2000) and Wandel et al. (1999). More and more evidence 
indicates that the assumption of virialization combined with 
the empirical relation of $R_{BLR} - L_{5100\AA}$ is a better method to 
estimate the BH masses of AGN (Ovcharov et al. 2005, Wu et al. 2004, 
McLure \& Jarvis 2004, 2002, Marziani et al. 2003), 
although there are few inconsistentcies of the reverberation Mapping technique 
(Maoz 1996). Recently, the method was used to estimate the BH masses for 
high redshift AGN (Dietrich \& Hamann 2004; Brotherton \& Scoggins 2004; 
McLure \& Jarvis 2002). However, there is an answered question : 
can the empirical relation between $R_{BLR}$ and continuum 
luminosity $L_{5100\AA}$ can be applied to any type of AGN. 
Wang \& Zhang (2003) have found that 
the size of BLRs of dwarf Active Galaxies (luminosity of H$\alpha$ 
less than $10^{41} {\rm erg\cdot s^{-1}}$, Ho et al. 1997a, 1997b, 
Maoz 1999) is not 
consistent with the value from the empirical relation 
$R_{BLR}\sim L_{5100\AA}^\alpha$ (new results about dwarf AGN 
can be found in Zhang, Dultzin-Hacyan \& Wang 2007a). 
Perhaps, we need more tests to 
verify whether the empirical relation can be applied to some special 
types of AGN, for which the sizes of BLRs can be estimated by other methods. 

   There is a special kind of AGN with DouBle-Peaked broad 
low-ionization emission lines (hereafter, dbp emitters) originated from the 
accretion disk near the central black hole. The most famous dbp emitters are 
NGC 1097 (Storchi-Bergmann et al. 1993, 1995, 1997, 2003), 
Arp102B (Chen et al. 1989, 1997, Chen \& Halpern  1989, 
Halpern et al. 1996, Antonucci et al. 1996, Sulentic et al. 1990) and 
3C390.3 (Shapovalova et al. 2001, Gilbert et al. 1999) etc.. A  
successful model for dbp emitters is the accretion disk model (Circular 
disk model: Chen et al. 1989, Chen \& Halpern 1989, Elliptical disk model: 
Eracleous et al. 1995, Warped disk model: Bachev 1999, 
Hartnoll \& Blackman 2000, Circular disk with spiral arm model: 
Hartnoll \& Blackman 2002, Karas et al. 2001). There are now two 
samples of dbp emitters, one includes 23 objects 
which are nearly all LINERs selected from radio galaxies (Eracleous \& Halpern 
1994, 2003, Eracleous et al. 1995), the other sample includes 112 
objects, of which 12\% are classfied as 
LINERs (Strateva et al. 2003) selected from SDSS DR2 (York et al. 2000). The 
statistical results indicate that the continuum luminosity of dbp emitters 
is not much different from the other AGN, the medium luminosity is 
$\sim 10^{44}{\rm erg\cdot s^{-1}}$. However, the line width is six times 
broader than that of normal AGN. Thus, if the BH masses of dbp emitters 
are estimated under the assumption of virialization and using the 
empirical relation between 
the size of BLRs and continuum luminosity, the BH masses  
would be about tens of times larger than those of normal AGN. 

   Fortunately, when fitting the line profiles of double-peaked broad 
emission lines of dbp emitters, the size of the region which produces the 
double-peaked broad emission lines can be estimated in units of 
$R_G=GM_{BH}/C^2$. This provides a better way 
to verify whether the size of BLRs of dbp emitters from the empirical 
relation $R_{BLR} - L_{5100\AA}^{\alpha}$ is the same as that from the 
accretion disk model, once the BH masses of dbp emitters are known. Section 2 
presents the data sample. The results are shown in Section 3, and then in 
Section 4 we present the discussions and conclusions. The cosmological 
parameters $H_{0}=75~{\rm km~s}^{-1}{\rm Mpc}^{-1}$, 
$\Omega_{\lambda}=0.7$ and $\Omega_{m}=0.3$ have been adopted here.

\section{Sample}

   We selected dbp emitters from the literature according to the 
following criteria. First, there is stellar velocity dispersion of the  
bulge or BH masses determined by kinematic/dynamical studies 
for the object, or there are other parameters, such as the accurate line 
width of narrow emission lines, which can trace the stellar velocity 
dispersion of bulge. Second, there are physical parameters of the 
accretion disk model for the double-peaked broad Balmer emission lines. 
We selected 12 dbp emitters listed in Table 1. 
There are four objects of which the spectra can be found in SDSS database:
B2 0742+31, CSO 0643, CBS 74 and 3C303. In SDSS, they are 
SDSS J074541.67+314256.7, SDSS J142314.19+505537.4, SDSS J083225.34+370736.2 
and SDSS J144302.76+520137.2. Thus, the line width of narrow 
emission lines can be measured accurately from SDSS spectra. 
In Table 1, the first 
column gives the name of the object, the second column gives the stellar
velocity dispersion in units of ${\rm km\cdot s^{-1}}$, the third column 
gives the logarithmic BH masses in units of ${\rm M_{\odot}}$ 
from the empirical relation $M_{BH}\sim\sigma^{4.02}$, the forth column gives the 
size of BLRs in units of light-days from
accretion disk model, the fifth column gives the continuum luminosity at
5100$\AA$ in units of ${\rm erg\cdot s^{-1}}$, the sixth column gives the 
line width of broad emission lines in units of ${\rm km\cdot s^{-1}}$, 
the seventh column gives 
the size of BLRs in units of light-days from the empirical relation
$R_{BLR} - L_{5100\AA}^{\alpha}$, the eighth column is
the logarithmic virial BH masses in units of $M_{\odot}$ estimated under 
the assumption of virialization,
the ninth column presents the power law index of the line emissivity for
each object according to accretion disk model, the last column
lists the references.  

\section{Results}
\subsection{Black Hole Masses}
  The most accurate way to estimate central BH masses is by kinematic 
or dynamics analysis. Using the results of some tens of nearby galaxies, 
a strongly 
tight relation between stellar velocity dispersion $\sigma$ and BH masses 
$M_{BH}(\sigma)$ has been found (Tremaine et al. 
2002, Ferrarese \& Merritt 2001, Gebhardt et al. 2000a): 
\begin{equation}
M_{BH}(\sigma) = 10^{8.13\pm0.06}\times(\frac{\sigma}{200{\rm km\cdot s^{-1}}})^{4.02\pm0.32} {\rm M_{\odot}}
\end{equation}
which indicates a tight relation between the evolution of bulge and that of 
the central black hole in galaxies, which is further confirmed 
by the tight relation between central BH masses and the masses of bulge of 
the host galaxy(H\"{a}ring \& Rix 2004, Marconi \& Hunt 2003, McLure \& 
Dunlop 2002, Laor 2001, Kormendy 2001, Wandel 1999). Whether the relation 
$M_{BH}\sim \sigma^{4.02}$ can be applied for higher redshift, higher 
luminousity AGN or other types of AGN has been studied by Treu et al. (2004), 
Nelson et al. (2004), Treves et al. (2003), Falomo et al. (2002), Wang \& Lu  
(2001) among others. These studies reach a similar conclusion 
that the empirical relation $M_{BH}\sim\sigma^{\alpha}$ holds for all AGN. 
Thus, we first estimate the central BH masses using this relation. There are eight 
dbp emitters with stellar velocity dispersions measured from  
absorption properties.
The other four dbp emitters are classified as QSOs in SDSS. 
It is difficult to measure the stellar velocity dispersion for QSOs. 
There are some other ways to estimate the central 
BH masses by physical parameters of bulge of the host galaxy, such as using 
the magnitude of the bulge (Mclure \& Dunlop 2001, 2002), 
using the line width of narrow emission lines (Borson 2003, Greene \& Ho 
2005a). Here, we measure the line width of the core of [OIII] doublet to trace 
the stellar velocity dispersion of the bulge. There can be also a 
broad gaussian function for the extended asymmetric components of 
the [OIII] emission lines as described in Greene \& Ho (2005a).  
The correlation between stellar velocity dispersion and line width of narrow 
emission lines has been studied by Greene \& Ho (2005a) for a large sample 
of AGN selected from SDSS. For radio loud AGN, 
the line width of [OIII]$\lambda5007\AA$ is broader than that of other narrow 
emission lines, however the line width of the core of 
[OIII]$\lambda5007\AA$ can trace stellar velocity dispersion. 
Further, the line width of 
the core of [OIII]$\lambda5007\AA$ of the four objects in our sample 
is the same as that of the other narrow emission lines within the errors. 
Here we assume the uncertainty of the line width to trace the stellar velocity 
dispersion is about 40 percent (the value for high radio AGN in Greene \& Ho 
2005a).
Once we obtain the central BH masses, the sizes of BLRs, $R_{BLR}(M)$, from 
accretion disk model can be translated from units of $R_G$ to units 
of light-days in order to compare with the values, $R_{BLR}(E)$, 
from the empirical relation $R_{BLR}(E) - L_{5100\AA}^{\alpha}$. 

   Furthermore, in order to compare the central BH masses estimated by 
different methods, the virial BH masses $M_{BH}(V)$ estimated under 
the assumption of virialization are also shown in Table 1:
\begin{subequations}
\begin{align}
&\frac{R_{BLR}(E)}{{\rm light-days}}=22.3\times(\frac{L_{5100\AA}}{10^{44}{\rm erg\cdot s^{-1}}})^{0.69}\\
&M_{BH}(V)=f_{FWHM}\times\frac{R_{BLR}(E)\times FWHM_{B}^2}{G}
\end{align}
\end{subequations}
where $FWHM_B$ is the line width of broad H$\beta$, $f_{FWHM}$ is a scale 
factor which depends on the structure of BLRs.
Here, we use the up-to-date empirical relation $R_{BLR} - L_{5100\AA}^{\alpha}$ 
selected from Kaspi et al. (2005). However, we notice that there are several  
dbp emitters for which the continuum luminosities are less than 
$10^{42}{\rm erg\cdot s^{-1}}$, which is out of the range of continuum 
luminosity of the sample of Kaspi et al. (2005). 
However, we have found that for low luminous AGN, the size of BLRs is 
somewhat larger than the predicted value from 
the empirical relation equation 2.a 
(Wang \& Zhang 2003; Zhang, Dultzin-Hacyan \& Wang 2007a). 
Thus, the BH masses estimated from the above equations 
for objects with lower continuum luminosity are smaller  
than the expected virial BH masses.

   Moreover, Collin et al. (2006) demonstrate that 
the scale factor $f$ depends on the line width of the broad emission lines, and 
this factor is much different from objects in Population A and Population B 
(Sulentic, Marziani \& Dultzin-Hacyan 2000) 
and the inclination effects play a role in some cases. 
We can not get an accurate spectra for all the dbp emitters to 
determine the factor of $f$.
Here, we use the mean facotr $f_{\sigma}=5.5$ for the second moment of 
broad emission lines rather than the value of FWHM (Onken et al. 
2004; Peterson et al. 2004), 
$M_{BH}=f_{\sigma}\frac{R_{BLR}\sigma_{B}^2}{G}$. 
We do not have 
enough information to get the mean/rms line spectra to determine the 
relation between FWHM(H$\beta_B$) and the second moment 
$\sigma_{H\beta_B}$, we commonly select the mean value of 1.87, 
$FWHM\sim1.87\sigma$, of the famous dbp emitter 3C390.3 (the line 
widths and second moment of the emission lines can be found in 
Peterson et al. 2004). Last, we accept $f_{FWHM}\sim1.56$ in equation 2b.
Also, the values of $R_{BLR}(E)$ from equation 2.a are shown 
in Table 1. Moreover, the continuum luminosity is the value after the 
subtraction of contributions of star light according to the starlight 
fraction in Eracelous \& Halpern (1994, 2003). For the two dbp emitters, 
NGC 1097 and NGC 4450, not included in the paper of Eracleous \& Halpern 
(1994, 2003), the continuum luminosities of the nucleus are 
estimated according to the absolute B magnitude of the featureless 
continuum of the nucleus from the paper of Ho et al. (2000). 

  Which parameters of double-peaked emission lines, FWHM, the 
second moment of the lines or other parameters, can be used to trace 
more accurately the velocity dispersion of BLRs is an open question. 
However, the main objective of this paper is to inspect whether the virial 
BH masses estimated by the line parameter FWHM are reliable. Thus 
the parameter, FWHM, is also selected from the paper of Strateva 
et al. (2003) and Eracleous \& Halpern (1994, 2003) measured 
according to the definition of FWHM: the full 
width at half maximum. Another problem is  the inclination angle 
of the accretion disk.
From the accretion disk model, we can accept the inclination angle of
accretion disk $i$, thus, the local value of FWHM in the accretion disk can
be similarly estimated by $FWHM_{local}\sim FWHM_{obs}/sin(i)$, which is
several times larger than the observed value of FWHM. However, the scale
factor $f_{FWHM}$ in equation 2.b is a factor to correct the effects of
inclination angle to some extent. Assuming a simple disk structure of BLRs,
the scale factor we used represents a mean inclination angle $i\sim30\degr$
(Onken et al., 2004). The inclination angles of dbp emitters in our sample
are always in the range from $\sim20\degr - \sim45\degr$
(Eracleous \& Halpern 1994, 2003), thus the inclination angle has little
effects on the results about BH masses. 

  Figure 1 shows the correlation between two kinds of BH masses: 
$M_{BH}(\sigma)$ and $M_{BH}(V)$. 
The Spearman rank correlation analysis gives the rank correlation coefficient 
($r_s$) of -0.23 and the significance level of its deviation from zero of 
$P_{null}\sim47\%$ for all 12 dbp emitters. The Kendalls correlation 
analysis presents the same results: correlation coefficient is -0.15 with 
$P_{null}\sim49\%$.  This result indicates that 
there is no significant correlation between the central BH masses estimated by 
means of the two different methods for dbp emitters. The mean value of the 
BH masses ratio of $M_{BH}(V)$ to $M_{BH}(\sigma)$ is no less than $30.47\pm13.83$,  
because the virial BH masses for two dbp emitters, NGC 1097 and NGC4050 
are smaller ones. If we consider only the six objects 
with measured stellar velocity dispersions of the bulge and with  
continuum luminosity larger than $10^{42}{\rm erg\cdot s^{-1}}$, the 
Spearman and Kendalls rank correlation coefficients are 0.48 with 
$P_{null}\sim33\%$ and 0.33 with $P_{null}\sim35\%$. Moreover, 
the Kolmogorov-Smirnov statistic analysis indicates that there are about 
32\% probability that the two kinds of BH masses having 
the same distribution. The ratio of $M_{BH}(V)$ to $M_{BH}(\sigma)$ for 
the six objects is $5.81\pm2.01$.

\subsection{The size of Broad Emission Line Regions}

  Now, the sizes of BLRs of dbp emitters can be calculated from two different 
methods: from the accretion disk model, $R_{BLR}(M)$, and from the empirical 
relation according to the continuum luminosity, $R_{BLR}(E)$. Here, the range 
of the size from accretion disk model is larger, so the flux-weighted mean 
radius is used as the mean size of BLRs (here, we can not consider the 
effects of eccentricity of elliptical accretion disk model):
\begin{equation}
\bar{R_{BLR}(M)} = \sum_{i}\frac{r_if_i}{\sum f_i}
\end{equation}
where $f_i=f_0\times r^{-q}$ is the line emissivity, the value of $q$ for each 
dbp emitter is listed in Table 1 according to the accretion disk model. Here, we 
just calculate the flux-weighted mean size by only even ten points between the 
inner radius and outer radius. Because of the steep power law 
of line emissivity, 
it is a better choice to select the points in logarithmic space of the 
radius. The last results are listed in Table 1. 
For object NGC 1097, we select the mean value of $r_q$ (inner the 
radius, the power-law index is $q=-1$ and outer the radius $q=1$) as the mean 
size of BLRs according to the fitted results for spectra observed during ten 
years (Storchi-Bergmann, Nemmen, et al. 2003; Storchi-Bergmann,
Eracleous et al. 1997; Storchi-Bergmann, Eracleous \& Halpern 1995;
Storchi-Bergmann, Baldwin et al. 1993). 

   The correlation between $R_{BLR}(E)$ and $R_{BLR}(M)$ is shown in 
Figure 2. The Spearman and Kendalls rank correlation coefficients are -0.31 
with $P_{null}\sim34\%$ and -0.24 with $P_{null}\sim27\%$ resepectively 
for all 12 dbp emitters. If we just consider the six dbp 
emitters with accurate stellar velocity dispersions and with continuum 
luminosities larger than $10^{42}{\rm erg\cdot s^{-1}}$, 
the Spearman and Kendalls rank correlation coefficients are 0.37 with 
$P_{null}\sim47\%$ and 0.21 with $P_{null}\sim57\%$ resepectively.
Furthermore, we notice that the size of BLRs according to the accretion 
disk model for 3C 390.3 is about 14.26 light-days as the same as 
the result $R_{BLR}\sim22.9_{-8.0}^{+6.3}$ from Reverberation Mapping 
Technique within the errors (Peterson et al., 2004). The mean ratio 
of $R_{BLR}(E)$ to $R_{BLR}(M)$ is about $50.44\pm34.91$ for all 12 dbp 
emitters and $2.48\pm1.22$ for the six dbp emitters with stellar veocity 
dispersions and with continuum luminosity larger than 
$10^{42}{\rm erg\cdot s^{-1}}$. Moreover, the Kolmogorov-Smirnov statistic 
analysis indicates that there are about 31\% probability that the two 
kinds of the size of BLRs have the same distribution.

  From the results about BH masses and size of BLRs, the basic results
are that the virial BH masses of dbp emitters are systematically 
larger than the BH masses 
estimated from the stellar velocity dispersions and the sizes of BLRs 
calculated from continuum luminosity are also systematically 
larger than the true sizes of 
BLRs from the accretion disk model. However, if only considering the 
six objects with stellar velocity dispersions and higher continuum 
luminosities, the results indicate that there are some moderate correlations 
between the two kinds of BH masses and the two kinds of the sizes of BLRs.

\section{Discussions and Conclusions}

   There are many references in the literature which discuss the 
consistency between the 
two kinds of BH masses according to the physical parameters of BLRs and 
according to the physical parameters of the bulge of the host 
galaxy, especially for low redshift and low luminosity AGN. It is not clear 
if the empirical relations (equation 2a and equation 2b) hold for AGN at $z>1$ 
or if the BH-Bulge relation holds for AGN with higher redshift 
(Sultntic et al. 2006). For dbp emitters, the double-peaked 
low-ionization emission lines can be best fitted by accretion disk 
model (Eracleous et al. 1995), the local velocity in the accretion 
disk can be estimated by Kepler's law, even considering the gravitational 
effects, especially when the radius is not nearer to the central black hole. 
Thus, the assumption of Virialization can be expected to hold for dbp 
emitters. However, we find that the mean virial BH masses are larger than 
BH masses from stellar velocity dispersion for dbp emitters. Although, we 
used the observed FWHM of broad emission lines, the scale factor 
$f_{FWHM}$ has considered the effects of inclination angle. Thus, the main 
reason of the difference of two kinds of BH masses is due to the larger 
size of BLRs derived from empirical relation $R\propto L_{5100\AA}^{\alpha}$.

  From the results above, BH masses of dbp emitters estimated from 
pure stellar velocity dispersion are about six times smaller than the 
virial BH 
masses. The BH masses estimated from line width of Narrow emission lines 
are about several tens of times smaller than the virial BH masses. 
Although, the slight correlation between BH masses $M_{BH}(\sigma)$ 
and virial BH masses $M_{BH}(V)$ can be found for dbp emitters with 
stellar velocity dispersions, we can not confirm the result due to 
the small number of our sample. Moreover, whether the line width of narrow 
emission lines can be used as the tracer of stellar velocity 
by $\sigma\sim 1\times\sigma_{line}$ for dbp emitters should be studied 
in the future using a large sample.

  More accurate estimation of BH masses for dbp emitter is necessary to 
estimate the fundamental physical parameter accretion rate $\dot{m}$. 
Using the BH masses from equations 2a and 2b, the dimensionless 
accretion rate $\dot{m}$ is several times smaller than that 
based on the BH masses from $M_{BH} - \sigma$. The mean dimensionless 
accretion rate $\dot{m}$ is about 0.01 from the analysed results 
of 135 dbp emitters by BH masses estimated under the assumption 
of virialization (Wu \& Liu, 2004, in their paper the BH masses 
are estimated from equation 2a and 2b). If we accept the BH masses 
ratio in our sample, the mean accretion rate $\dot{m}$ based on the BH masses 
from stellar velocity dispersions should be 
about 0.1. So, the ADAF accretion flow ($\dot{m}\le0.28\times\alpha^2$ 
and $\alpha\sim0.1 - 0.3$, Mahadevan, 1997; Mahadevan \& Quataert, 1997; 
Narayan et al., 1995) should just exist in a much smaller part of dbp emitters 
according to the value of accretion rate.

   We have shown that the relation between the size of BLRs and 
continuum luminosity does not hold for dbp emitters. Also, 
BH masses estimated from line width of narrow emission lines 
have large errors. The mean 
flux-weighted size of BLRs is also several times smaller than the value 
estimated from the empirical relation $R_{BLR} - L_{5100\AA}^{\alpha}$.
This is perhaps due to the following reasons according to the definition 
of ionization parameter:
\begin{equation}
\Gamma = \frac{Q}{4\pi r^{2}cN_e}
\end{equation}
The first possible reason is due to the different value of $Q$ between dbp 
emitters and normal AGN. However, as we discussed in another paper 
(Zhang, Dultzin-Hacyan \& Wang 2007b), there is the same strong 
correlation between the luminosity of H$\alpha$ and 
continumm luminosity $L_{5100\AA}$ for dbp emitters as that 
for normal AGN (Greene \& Ho 2005b). Thus, there are not much different 
effects of ionization continuum on the size of BLRs for dbp emitters 
from those for normal AGN. The second possible reason is due to 
higher electron density of BLRs of dbp emitters than that of normal AGN. 
More and more evidence indcates that double-peaked broad emission lines 
originate from the accretion disk near the central black hole as mentioned 
in the introduction. This high electron density region is also responsible 
for broad FeII emission lines observed in many AGN (Ferland \& Person 1989).
The third possible reason is 
due to the different ionization parameter of BLRs of dbp emitters. 
We think the higher electron density of BLRs 
of dbp emitters is the dominated reasone which leads to smaller size 
of BLRs.
 
  The last summary is: First: for dbp emitters, the BH masses 
estimated under the assumption of virialization using the continumm 
luminosity and line width of broad H$\beta$ are much larger than the 
BH masses from $M_{BH} - \sigma$ which is a more accurate relation. 
Second, the sizes of the BLRs of dbp emitters can not be estimated by the 
empirical relation of $R_{BLR} - L_{5100\AA}$. Sizes estimated from this relation 
are larger than the mean flux-weighted sizes of BLRs of dbp emitters.
The higher electron density of BLRs of dbp emitters would be the 
main reason which leads to smaller size of BLRs than that of normal AGN.
The dimensionless 
accretion rate, one of the fundamental parameters for AGN, depends 
sensitively on the central BH masses, thus, to find a more accurate and 
convenient way to estimate the BH masses for dbp emitters is an important 
topic for the future.

\section*{Acknowledgements}
We are grateful to the anonymous referee for helpful suggestions which 
improve and clarify our paper.
ZXG gratefully acknowleges the postdoctoral scholarships offered by la
Universidad Nacional Autonoma de Mexico (UNAM). D. D-H acknowledges 
support from grant IN100703 from DGAPA, UNAM. This research has made use 
of the NASA/IPAC Extragalactic Database (NED) which is operated by the Jet 
Propulsion Laboratory, California Institute of Technology, under contract 
with the National Aeronautics and Space Administration. This research has 
also made use of the HyperLeda.
This paper has also made use of the data from the SDSS projects.
Funding for the creation and the distribution of the SDSS Archive
has been provided by the Alfred P. Sloan Foundation, the
Participating Institutions, the National Aeronautics and
Space Administration, the National Science Foundation,
the U.S. Department of Energy, the Japanese Monbukagakusho, and the
Max Planck Society. The SDSS is managed by the Astrophysical Research
Consortium (ARC) for the Participating Institutions. The Participating
Institutions are The University of Chicago, Fermilab, the
Institute for Advanced Study, the Japan Participation Group,
The Johns Hopkins University, Los Alamos National Laboratory,
the Max-Planck-Institute for Astronomy (MPIA),
the Max-Planck-Institute for Astrophysics (MPA), New
Mexico State University, Princeton University, the United
States Naval Observatory, and the University of Washington.

\begin{table*}
\centering
\begin{minipage}{170mm}
\caption{Data of Sample}
\begin{tabular}{llllllllll}
\hline
Name & $\sigma$ &  $\log(M_{\bullet}(\sigma))$ & 
$R_{BLR}(M)$ & $L_{5100\AA}$  & FWHM & $R_{BLR}(E)$ 
& $\log(M_{\bullet}(E))$ & $q$ & Ref. \\
\hline
NGC 1097 &  208$\pm$5$^a$ & 8.19$\pm$0.10 & 10.29 & 41.3 & 13700 & 0.28L & 
7.22 & $\star$ & 1,2,3,4 \\
3C390.3 & 273$\pm$16$^a$ & 8.67$\pm$0.21 & 14.26 & 43.87 & 11900 & 18.08 & 
8.89 & 3.0 & 2,5\\
Pictor A & 145$\pm$20$^a$ & 7.57$\pm$0.26 & 2.15 & 43.2 & 18400 & 6.14 & 
8.81 & 1.5 & 2,6,7\\
Arp 102B & 188$\pm$8$^a$ & 8.02$\pm$0.12 & 2.91 & 42.48 & 16000 & 1.93 & 
8.19 & 3.0 & 2,4,5\\
PKS 0921-213 & 144$\pm$17$^a$ & 7.56$\pm$0.22 & 3.41 & 43.06 & 8300 & 
4.91 & 8.02 & 1.5 & 2,6\\ 
IE 0450.3-1817 & 150$\pm$26$^a$ & 7.63$\pm$0.32 & 3.05 & 42.07 & 10900 & 
0.99 & 7.57 & 3.0 & 2,5\\
IRAS 0236.6-3101 & 154$\pm$15$^a$ & 7.67$\pm$0.18 & 3.29 & 44.13 & 7800 & 
27.49 & 8.72 & 3.0 & 2,5\\ 
B2 0742+31 & 138.5$^b$ & 7.49$\pm$0.74 & 1.38 & 45.46 & 6500 & 
234.58 & 9.49 & 2.0 & 3\\
CSO 0643 & 116.9$^b$ & 7.19$\pm$0.71 & 2.01 & 44.58 & 9000 & 
56.79 & 9.15 & 3.0 & 6\\
CBS 74 & 87.3$^b$ & 6.68$\pm$0.67 & 0.06 & 44.07 & 9200 & 24.96 & 
8.82 & 3.0 & 6\\
3C 303 & 115.9$^b$ & 7.18$\pm$0.71 & 1.17 & 43.39 & 6800 & 
8.34 & 8.08 & 3.0 & 5\\
NGC 4450 & 140$^c$ & 7.51 & 1.96 & 40.65 & 9500 & 0.11L & 6.45 & 3.0 & 
4,8\\
\hline
\end{tabular}
\\
a represents that the stellar velocity dispersion is measured from the 
absorption spectra. \\
b represents that the stellar velocity dispersion is estimated from the 
line width of narrow emission lines and with 40 percent uncertainty.\\
c represents that the stellar velocity is selected from HYPERLEDA 
(http://www-obs.univ-lyon1.fr/hypercat/).\\
$\star$ represents the mean radius with 1108$R_G$\\
The first column is the name of the object, the second column is the stellar 
velocity dispersion in units of ${\rm km\cdot s^{-1}}$, the third column is the 
logarithmic BH masses in units of ${\rm M_{\odot}}$ from $M_{BH} - \sigma$ 
relation, the forth column is the size of BLRs in units of light-day from 
accretion disk model, the fifth column is the continuum luminosity at 
5100$\AA$ in units of ${\rm erg\cdot s^{-1}}$ after the subtraction of contribution 
of star light, the sixth column is the line width 
of broad emission lines in units of ${\rm km\cdot s^{-1}}$, the seventh column is 
the sizes of BLRs in units of light-day from the empirical relation 
$R_{BLR} - L_{5100\AA}^{\alpha}$ ('L' means that the size of BLRs 
should be larger than 
the listed value), the eighth column is 
the  logarithmic BH masses in units of ${\rm M_{\odot}}$ estimated under 
the assumption of virialization, 
the ninth column presents the power law index of the line emissivity for 
each object according to the accretion disk model, the tenth column 
presents the references. \\
The references are:\\
1: Storchi-Bergmann, Nemmen, et al. 2003; 2: Lewis \& Eraclous 2006; 
3: Eracleous et al. 1995; 4: Ho et al. 2000; 5: Eracleous \& Halpern 1994; 
6: Eracleous \& Halpern 2003; 7: Sulentic et al. 1995; 
8: Fillmore et al. 1986
\end{minipage}
\end{table*}

\begin{figure}
\includegraphics[width=82mm]{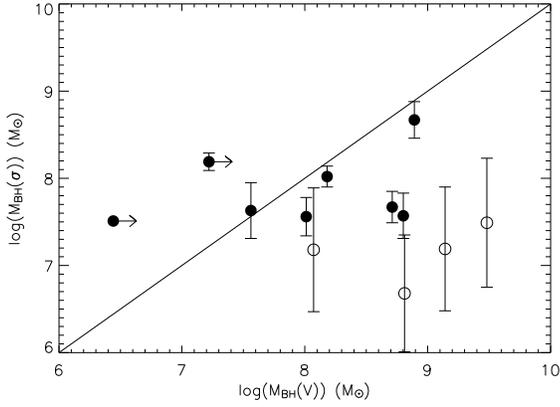}
\caption{The correlation between $M_{BH}(\sigma)$ and $M_{BH}(V)$. The solid 
circle represents the object of which $M_{BH}(\sigma)$ is estimated by the 
stellar velocity dispersion $\sigma$, the open circle represents the object 
of which $M_{BH}(\sigma)$ is estimated by the line width of narrow emission 
lines. The arrow represents that the value of $M_{BH}(V)$ is the lower 
limit one.
The solid line represents the relation $M_{BH}(\sigma) = M_{BH}(V)$.}
\label{fig1}
\end{figure}

\begin{figure}
\includegraphics[width=82mm]{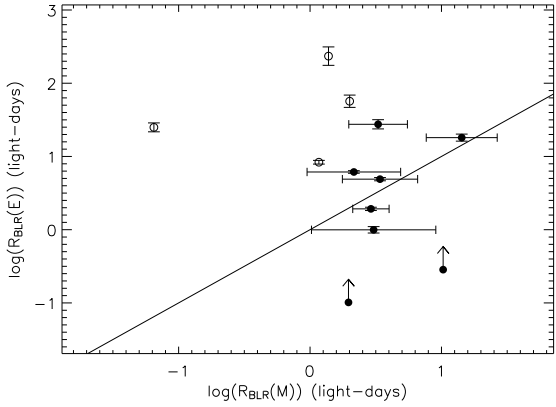}
\caption{The correlation between $R_{BLR}(M)$ and $R_{BLR}(E)$. The solid
circle represents the object of which $M_{BH}(\sigma)$ is estimated by the
stellar velocity dispersion $\sigma$, the open circle represents the object
of which $M_{BH}(\sigma)$ is estimated by the line width of narrow emission
lines. The arrow represents that the value of $R_{BLR}(E)$ is a lower 
limit one. 
The uncertainty of $R_{BLR}(E)$ is given by assuming 5 percent 
uncertainty in continuum luminosity. 
The solid line represents the relation $R_{BLR}(M) = R_{BLR}(V)$.}
\end{figure}

\label{lastpage}

\end{document}